\documentclass[aps,prl,preprintnumbers,groupedaddress,nofootinbib,showkeys,showpacs,twocolumn]{revtex4}

\usepackage{amsfonts}
\usepackage{mathrsfs}
\usepackage{graphicx}
\usepackage{picins}
\usepackage[colorlinks=true]{hyperref}
\usepackage{color}

\newcommand{\be}{\begin{equation}}
\newcommand{\en}{\end{equation}}
\newcommand{\bea}{\begin{eqnarray}}
\newcommand{\ena}{\end{eqnarray}}

\def\to{\rightarrow}
\def\vec{\mathbf}

\def\etal{{\it et. al.}}

\def\d{{\rm d}}

\def\paslash{\partial\!\!\!\slash }

\def\sslash{s\!\!\!\slash }

\begin{document}

\title{The Aharonov-Casher and scalar Aharonov-Bohm topological effects}

\author{Sayipjamal Dulat}
\email{dulat98@yahoo.com}

\author{Kai Ma}
\email{MakaiNCA@gmail.com}

\affiliation{School of Physics Science and Technology, Xinjiang
University, Urumqi, 830046, P. R. China}

\begin{abstract}

\noindent We reexamine the topological and nonlocal natures of the
Aharonov-Casher and scalar Aharonov-Bohm phase effects. The
underlying $U(1)$ gauge structure is exhibited explicitly. And the
conditions for developing topological Aharonov-Casher and scalar
Aharonov-Bohm phases are clarified.
We analyse the arguments of M.
Peshkin and H. J. Lipkin (Phys. Rev. Lett. 74, 2847(1995)) in detail
and show that they are based on the wrong Hamiltonian which yields
their conclusion  incorrect.

\end{abstract}

\keywords{Topological phase, gauge symmetry }

\pacs{03.65.Vf, 11.15.-q, 11.30-j}

\maketitle


\textbf{\emph{Introduction}}---In 1959 Aharonov and Bohm
(AB)\cite{AB} proposed two types of electron interference
experiments the usual magnetic (or
vector) AB effect, and the less often mentioned electric (scalar) AB
effect to exhibit the significance of electromagnetic
potentials in quantum mechanics. The essential physical aspect of the usual magnetic (or
vector) AB experiment\cite{ABE1, ABE2} is that electrons passing
around opposite sides of a long solenoid acquire a phase shift due
to the enclosed magnetic flux even though the electrons move in a
region free of electric and magnetic fields but nonzero magnetic
potentials.  The electric (or scalar) AB effect for charged
particles concerns the phase shift caused by the scalar potential
$V=-eU$ in the Schroedinger equation. However, the experiment has
not yet been performed because of technical difficulties with
electron interferometers. Rather than using electrons acted on by
electrostatic potentials, Ref. \cite{SAB1, SAB2, SAB3} have
performed an analogous interferometry experiment with thermal
neutrons subject to pulsed magnetic fields. And phase shifts have
been observed  to a high degree of accuracy. In the neutron
interferometry experiment, the phase shift is due to a scalar
potential, $V=-\vec\mu\cdot\vec B$, which is the analog of $V=-eU$,
the scalar potential in the scalar $AB$ effect for electrons. In
1984 Aharonov and Casher(AC)\cite{AC} pointed out that in the
nonrelativistic limit the wave function of a neutral spin-$1/2$
particle with magnetic dipole moment $\mu$ develops a topological
phase when traveling in a closed path which encircles an infinitely
long filament carrying a uniform charge density. This effect has
also been observed experimentally \cite{ACE1,ACE2}. The AC
experiment with neutrons is the analog of the magnetic (vector) AB
experiment for electrons. In 1987 Boyer \cite{ACC1} argued that a
classical neutron does experience a force, causing a velocity change
which depends upon the side of the wire that the neutron passes. In
1988 Aharonov\etal~\cite{ACC2} elaborated furthermore the force free
nature of AC effect. In 1991 He and McKellar \cite{HM} provided the
conditions for developing a topological AC phase for a neutral spin
half particle in $2+1$  space time dimensions. In 1995 Peshkin and
Lipkin (PL) \cite{ACC3} argued that both SAB and AC effects are not
topological due to the torque action on the spin of neutron. And
extensive controversies \cite{ACC41} have followed about the
nonlocal force free nature of the SAB and AC effects. Thus, further
investigations about the gauge invariance principle for
topological SAB and AC phases of a neutral spin half particle are
necessary. In this letter we will address the underlying principle
for the generation of the topological SAB and AC phase effects.

\textbf{\emph{The underlying $U(1)$ gauge symmetry}}---Now we study how the gauge invariance principle determines the topological phases of the AC and SAB effects for a neutral spin half particle with an anomalous magnetic dipole
moment $\mu$. The Lagrangian
for a neutral spin half particle  with an anomalous magnetic dipole
moment $\mu$ interacting with the electromagnetic field has
the form,
\begin{equation}\label{CL}
    \mathcal {L}=\bar{\psi}(x)
    (i\paslash-m)\psi(x)-\frac{\mu
    }{2}\bar{\psi}(x) \sigma^{
    \mu\nu}F_{\mu\nu}\psi(x),
\end{equation}
where $m$ is the mass of the particle; the $\sigma^{\mu\nu}=i[\gamma^{\mu},\gamma^{\nu}]/2$
represents the internal spin structure of the neutral particle, and $1/2$
accounts for the antisymmetry both of $\sigma^{\mu\nu}$ and
$F_{\mu\nu}$.
The spin projection operator is
$$
\hat\Sigma_\pm(\vec s)=(\mathbb{I} \pm \gamma^5\sslash)/2 ,
$$
where $\vec s$ is the unit polarization vector in the rest frame $(\vec s\cdot \vec s =1)$; and the
four-spin vector $s^\mu$ in an arbitrary inertial frame which has the form,
$$ s^\nu=\bigg(\frac{\vec p \cdot \vec s} {m},
    \vec s + \frac{\vec p(\vec p \cdot\vec s)}{m(p^0 + m)}\bigg),
    $$
where $p^{\mu}=(p^0,\vec p)$ is the four momentum of the
particle. To separate spin-up and spin-down parts, we insert the
unite operator $\mathbb{I}=\hat \Sigma_+(\vec s)+\hat
 \Sigma_-(\vec s)$ into (\ref{CL}), and we have,
 $$
 \mathcal L=\mathcal {L}_+ + \mathcal {L}_-\;,
 $$
where
\begin{eqnarray}\label{separated-cl}
  \mathcal {L}_\pm &=& \bar{\psi}(i\paslash-m-\frac{\mu}{2}\sigma^{\mu\nu}F_{\mu\nu})
  {\hat\Sigma}_\pm(\vec s)\psi \nonumber\\
&=&\frac{1}{2}\bar{\psi}
    (i\paslash-m)\psi \pm \frac
    {\mu}{4}\bar{\psi}
    \gamma^{\alpha}\epsilon_{\alpha
    \beta\mu\nu}s^\beta F^{\mu
    \nu}\psi,
\end{eqnarray}
here, we have used the following relations,
\begin{equation}\label{MI1}
    \bar{\psi}(i\paslash-m)
    \hat\Sigma_\pm({\vec s})\psi=
    \frac{1}{2}\bar{\psi}
    (i\paslash-m)\psi,
\end{equation}
\begin{equation}\label{MI2}
    \frac{\mu}{2}\bar{\psi}
    \sigma^{\mu\nu}F_{\mu\nu}
    \hat\Sigma_\pm({\vec s})\psi=\mp\frac
    {\mu}{4}\bar{\psi}\gamma
    ^{\alpha}\epsilon_{\alpha
    \beta\mu\nu}s^\beta F^{\mu
    \nu}\psi.
\end{equation}
Explicitly, $\mathcal {L}_{+}$ and $\mathcal {L}_{-}$ are equivalent
except a minus sign. Now we define a new quantity $\mathcal {A}_{\alpha}(x)$ as,
\begin{equation}\label{NVP}
    \mathcal{A}_{\alpha}(x)=
    -\frac{1}{2}\epsilon_
    {\alpha\beta\mu\nu}s^{
    \beta}F^{\mu\nu}
    .
\end{equation}
Then the lagrangian (\ref{separated-cl}) can be written as,
\begin{equation}\label{NL}
    \mathcal {L}_\pm=
    \frac{1}{2}\bar{\psi}
    i\gamma^{\alpha}(\partial_{
    \alpha}\pm i\mu \mathcal {A}
    _{\alpha})\psi-\frac{m}
    {2}\bar{\psi}\psi,
\end{equation}
where the common factor $1/2$ is due to the spin
projection.
Under a new $U_{\rm mm}(1)$ local gauge transformation (mm stands for the
magnetic moment),
\begin{equation}
\mathcal {A}_{\alpha}(x)\to
    \mathcal {A}_{\alpha}
    (x)\mp\partial_{\alpha}\phi(x)\;,
\end{equation}
the gauge invariance of the theory  (\ref{NL}) requires
the wave function $\psi$ to transform as,
\begin{equation}\label{NGF}
    \psi(x) \to e^{\pm i\mu\phi(x)}
    \psi(x)\;.
\end{equation}
From the Lagrangian (\ref{NL}) (in the following we will consider only the positive part of (\ref{NL}).), the Euler-Lagrangian equation for $\psi$ is,
\begin{equation}\label{me}
    \big[\gamma^{\mu}(p_{\mu}-
    \mu\mathcal {A}_{\mu})
    -m\big]\psi(x)=0,
\end{equation}
the conserved magnetic dipole momemnt current density is,
\begin{equation}\label{conservation-equation}
    J^{\mu}_{\rm mm}=\mu\bar\psi\gamma^{\mu}\psi\;,\;\;\;\;\partial_{\mu}J^{\mu}_{\rm mm}=0~.
\end{equation}
Now let us consider that our particle always moves in the space-time region where  effective electromagnetic field strength $\mathcal{F}_{\alpha\beta}=\partial_{\alpha}\mathcal{A}_{\beta}- \partial_{\beta}\mathcal{A}_{\alpha}=0$ but $\mathcal{A}_{\alpha}\neq0$. Since $\mathcal{F}_{\alpha\beta}$ is antisymmetric, $\mathcal{A}_{\alpha}$ can be written as a 4-dimensional gradient of a scalar function $\phi(x)$,
\begin{equation}\label{GC}
    \mathcal {A}_{\alpha}
    (x)=\partial_{\alpha}\phi(x).
\end{equation}
In this case, $\mathcal{A}_{\alpha}$ is a pure vector gauge potential, and the equation of motion for a force free particle can be written as,
\begin{equation}
    \big[\gamma^{\mu}p_{\mu}
    -m\big]\psi^\prime(x)=0.
\end{equation}
where the force free solution is,
\begin{equation}
\psi^\prime(x)=\exp{[i\mu\int_{x_{0}}^{x}\d z_{\mu}\mathcal {A}^{\mu}(z)]}\psi.
\end{equation}
To generate a nontrivial phase, the space-time region in which $\mathcal{F}_{\alpha\beta}=0$ but $\mathcal{A}_{\alpha}\neq0$ must have a singular point in it, such that $\mathcal{F}_{\alpha\beta}\neq0$ at this singular point. If one now transports the particle around a closed path in the $\mathcal{F}_{\alpha\beta}=0$ region which encloses regions with $\mathcal{F}_{\alpha\beta}\neq0$, the wave function develops a nontrivial phase difference,
\begin{equation}\label{total-phase-diff}
    \Delta\phi=\frac{\mu}{2}\int_{\Sigma}\d S^{\mu\nu}\mathcal{F}_{\mu\nu},
\end{equation}
where, $\Sigma$ is the surface enclosed by the trajectory of matter particle. For a non-closed path in the space-time region $\mathcal{F}_{\alpha\beta}=0$, the wave function develops a relativistic phase,
\begin{equation}\label{CACP}
    \phi=\mu\int_{r^{\beta}_{i}}
    ^{r^{\beta}_{f}}\d r_{\beta}
    \mathcal {A}_{\beta}=-\frac{
    \mu}{2}\int_{r^{\beta}_{i}}^
    {r^{ \beta}_{f}}\d r_{\beta}
    \varepsilon^{\beta \alpha\mu
    \nu}s_{\alpha}F_{\mu\nu} (t,
    \vec{r}).
\end{equation}
The ordinary AC phase can be recovered if there is only electric field $\vec{E}$ present in  $\mathcal{F}_{\alpha\beta}=0$ region,
\begin{widetext}
\begin{equation}\label{NRLACP}
    \phi_{\rm AC}=\mu
    \int_{\vec{r}_{i}}^{\vec{r}_{
    f}}\d r_{i}\varepsilon^{ijk0
    }s_{j}F_{k0}(t,\vec{r})
    =\mu\int_{\vec{r}_{i}}^
    {\vec{r}_{f}}\d\vec{r}\cdot
    \bigg[\bigg(\vec{s}+
    \frac{\vec{v}(\vec{p}\cdot
    \vec{s})}{(p^0+m)}
    \bigg)\times\vec{E}\bigg]
    =\mu\int_{\vec{r}_{i}}^
    {\vec{r}_{f}}\d\vec{r}\cdot
    \bigg(\vec{s}\times\vec{E}\bigg).
\end{equation}
The SAB phase that is related to a magnetic field $\vec B$ in $\mathcal{F}_{\alpha\beta}=0$ region has the following form,
\begin{equation}\label{NRLSABP}
    \phi_{\rm SAB}=-\frac
    {\mu}{2}\int_{t_{i}}^{t_{f}}\d
    r_{0}\varepsilon^{0ijk}s_{i} F
    _{jk}(t,\vec{r})+\frac{\mu}{2}
    \int_{\vec{r}_{i}}^{\vec{r}_{f}} \d r_{i}
    \varepsilon^{i0jk}s_{0} F_{jk}
    (t,\vec{r})=-\mu\int_{t_{i}}^{t
    _{f}}\d t(\vec{s}\cdot\vec{B})
    -\mu\int_{\vec{r}_{i}}^{\vec{r
    }_{f}}\d\vec{r}\cdot\vec{B}
    (\vec{v}\cdot\vec{s}),
\end{equation}
\end{widetext}
where the second term is the relativistic
correction. In the  SAB experiment, polarization vector $\vec s$
and velocity  vector $\vec{v}$ are constant;
the $\vec s$, $\vec{v}$ and $\vec{B}$ are all along the
same direction. Thus, when the trajectory of matter particle forms
a closed path, the additional contribution, the magnetic flux
through the surface surrounded by the loop, is zero. Of course,
the original AC and SAB phases can be recovered by  applying directly
the nonrelativistic polarization vector $s^{\alpha}\approx(0,\vec s)$.

\textbf{\emph{Conditions for AC and SAB phases}}
By taking the nonrelativistic limit, now we discuss the conditions for developing the topological AC and SAB phases.
To obtain the non-relativistic Hamiltonian, we can employ the Foldy-Wouthuysen
(WF)transformation \cite{FW}.
The relativistic Hamiltonian can be obtained from the equation of motion (\ref{me}),
\begin{equation}\label{SDH}
    H_{\hat\Sigma}=\vec{\alpha}\cdot(
    \vec{p}-\mu\vec{\mathcal {A}
    })+\beta m+\mu\mathcal {A}^{
    0}.
\end{equation}
Due to similarity with the Dirac Hamiltonian, one can easily obtain
the FW-transformed Hamiltonian,
\begin{widetext}
\begin{equation}\label{FWH}
    H_{\hat\Sigma-\rm FW}=\beta\bigg(m+
    \frac{1}{2m}(\vec{p}-\mu
    \vec{\mathcal {A}})^2
    \bigg)+\mu\mathcal
    {A}^{0}-\frac{\mu}{2m}\beta(
    \vec{\Sigma}\cdot\vec{
    \mathcal {B}})-\frac{i\mu}
    {8m^2}\vec{\Sigma}\cdot(\vec
    {\nabla}\times\vec{\mathcal {E}})
    -\frac{\mu}{4m^2}\vec{\Sigma}\cdot
    (\vec{\mathcal {E}\times\vec{p}})-
    \frac{\mu}{8m^2}\vec{\nabla}\cdot
    \vec{\mathcal {E}},
\end{equation}
\end{widetext}
where,
\begin{equation}\label{sigma}
    \beta=\left(
             \begin{array}{cc}
               \mathbb{I} & 0 \\
               0 & -\mathbb{I} \\
             \end{array}
           \right),~~~
    \vec\Sigma=\left(
             \begin{array}{cc}
               \vec\sigma & 0 \\
               0 & \vec\sigma \\
             \end{array}
           \right),~
\end{equation}
$\vec\sigma=(\sigma_{1},\sigma_{2},\sigma_{3})$ are Pauli matrixes,
and
\begin{eqnarray}
    &&\vec{\mathcal {A}}=\vec{s}
    \times\vec{E},~~{\mathcal {A
    }}_{0}=\vec{s}\cdot\vec{B},~
    \label{NRGP}\\
    &&\vec{\mathcal {B}}=\vec{
    \nabla}\times\vec{\mathcal
    {A}}=\vec{s}(\vec{\nabla}
    \cdot\vec{E})-(\vec{s}\cdot
    \vec{\nabla})\vec{E}
    ,~\label{EM}\\
    &&\vec{\mathcal {E}}=-\vec
    {\nabla}\cdot\mathcal {A}^
    {0}=-\vec{\nabla}(\vec{s}
    \cdot\vec{B})\label{EE}.
\end{eqnarray}
The  terms which induce the ordinary AC and SAB effects are
the first two terms in (\ref{FWH}). Thus, in order to generate a
pure phase effect, the additional terms in (\ref{FWH}) which affect
the energy  of this system should vanish,
namely,
\begin{equation}\label{GPCE}
    \vec{\mathcal {B}}=0,~~
    \vec{\mathcal {E}}=0.
\end{equation}
By using Eqs.
(\ref{EM}) and (\ref{EE}) we can obtain the following conditions to generate the AC and SAB phases, respectively,
\begin{eqnarray}
   &&\vec{\nabla}\cdot\vec{E}=0,
   ~~(\vec{s}\cdot\vec{\nabla} )
   \vec{E}=0; \label{PCFAC}\\
   &&\vec{\nabla}(\vec{s}\cdot
   \vec{B})=0, \label{PCFSAB}\
\end{eqnarray}
 thus no
force or torque acts on the particle. In the ordinary AC configuration, the
nontrivial topological spacetime region is generated by the condition
$\vec{\nabla}\cdot\vec{E}\neq0$.

\textbf{\emph{Remarks}}----
Ref. \cite{ACC3} analyzed measurable dynamical variables, and
the spin autocorrelation operators by using the following Hamiltonian
for the SAB effect,
\begin{equation}\label{PL}
    H={\vec p}^2/2m -\mu\;\vec\sigma\cdot\vec B(t)\;,
\end{equation}
where $\vec\mu$ is the neutron's magnetic moment, $\vec B$ is the
magnetic field.
The Heisenberg  equation of motion of dynamical variable $\vec{\sigma}$ is ,
\begin{equation}\label{motion-sigma}
    \frac{\hbar}{2}\dot{\vec{\sigma}}=\mu\vec{\sigma}\times\vec{B}
\end{equation}
and the spin autocorrelation operators are,
\begin{eqnarray}
    C(\Delta t)
    &=&\frac{1}{4}[\sigma_{x}(t_{f})\sigma_{x}(t_{i})+\sigma_{y}(t_{f})\sigma_{y}(t_{i})+{\rm H.C.}],\label{SACOC}\\
    S(\Delta t)
    &=&\frac{1}{4}[\sigma_{y}(t_{f})\sigma_{x}(t_{i})-\sigma_{x}(t_{f})\sigma_{y}(t_{i})+{\rm H.C.}]\label{SACOS}.
\end{eqnarray}
and claimed that both SAB and AC effects are not topological due to the torque action.
However, we have showed
in the above discussions that, when the polarization direction of neutron don't vary, the
underlying gauge symmetry is $U(1)_{\rm mm}$. Furthermore, by (\ref{FWH}) the precessing equation
for spin operator  is,
\begin{equation}\label{precessing-equation}
    \frac{\hbar}{2}\dot{\vec{\sigma}}
    \propto\vec{\sigma\times(\vec{\mathcal {B}}}+\vec{\mathcal {E}}\times\vec{v}).
\end{equation}
thus for a spin-$1/2$ neutral particle with fixed spin polarization, the correct term which describes the spin precessing around it's polarization direction is described by the
$\vec{\sigma}\cdot\vec{\mathcal {B}}$ in the SAB configuration and the $\vec{\sigma}\times\vec{\mathcal
{E}}$ in the AC configuration. If we make replacements $\vec{\mathcal {B}}\to\vec{B}$
and $\vec{\mathcal {E}}\to\vec{E}$, we can recover the Hamiltonian
employed by Peshkin and Lipkin \cite{ACC3}.
Furthermore, both in the AC and SAB cases, $\vec{\mathcal {B}}=0$ and $\vec{\mathcal {E}}=0$. Thus there are no dynamical variables available to describe the spin precession along it's polarization direction. This also can be realized from the correlation operators (\ref{SACOC}) and (\ref{SACOS}). In showing this, we firstly introduce two new correlation operators,
\begin{eqnarray}
  \mathcal{S}_{-}&=&\sigma_{+}(t_{f})\sigma_{-}(t_{i})~,\\
  \mathcal{S}_{+}&=&\sigma_{-}(t_{f})\sigma_{+}(t_{i})~.
\end{eqnarray}
The physical meaning of the operator $\mathcal{S}_{-}$ is that at time $t_i$ the ``spin up" state changes into  ``spin down" state, and at time $t_f$ the ``spin up" state recovers. Note that the ``spin up" and ``spin down" states at different time may have different reference direction, i.e., in general the spin polarization is time dependent $\vec{s}(t)$. In one word, these correlation operators describe the spin flip between the ``spin up" and ``spin down" polarization states at time interval $\Delta t=t_f-t_i$. By using relations $[\sigma_{a}(t'),\sigma_{b}(t)]=0$ ($a,b=x,y,z$) for $t'\neq t$, the spin autocorrelation operators (\ref{SACOC}) and (\ref{SACOS}) can be written as,
\begin{equation}\label{SACO}
    C(\Delta t)=\mathcal{S}_{+}+
    \mathcal{S}_{-},~~
    S(\Delta t)=i(\mathcal{S}_{+} -
    \mathcal{S}_{-}).
\end{equation}
Then $C(\Delta t)$ and $S(\Delta t)$ have the same physical meaning as the $\mathcal{S}_{-}$ and $\mathcal{S}_{+}$. However, such spin flip processes have not been observed experimentally yet. On the other hand, such processes may occur at the quantum fluctuation level, but such fluctuation exists only at a very short time interval, $\Delta t\approx0$.
One may note that $\Delta t=\tau$, where
the $\tau$ in Ref.\cite{ACC3} is the time for the particle in the magnetic field, it may be very long, $\tau\to\infty$. Thus the correlation operators (\ref{SACOC}) and (\ref{SACOS}) should vanish at time interval $\tau$.

\textbf{\emph{Conclusions}}---We have established  the topological and nonlocal nature of the AC and SAB phases on 4-dimensional space-time. The
underlying $U_{\rm mm}(1)$ gauge structure is exhibited explicitly.
On the other hand we have derived the corresponding conditions for
the generation of these phases from the underlying principle, and these conditions are consistent with the experimental phenomena both for AC and SAB effects. The
meaning of these conditions are clarified. The corresponding $U_{\rm mm}(1)$ gauge symmetry ensures the force free nature of the
induced topological AC and SAB phases.
The statement by M.
Peshkin and H. J. Lipkin \cite{ACC3}
that the AC and SAB effects are not topological effects is examined.
We analyse their arguments in detail and show that the correct terms which describe the
precessing of spin direction along the polarization direction are
$\vec{\sigma}\cdot\vec{\mathcal {B}}$ in the SAB configuration and $\vec{\sigma}\cdot(\vec{\mathcal
{E}}\times\vec{v})$ in the AC configuration. Thus they are based on
the wrong Hamiltonian which yields their incorrect conclusion.

\textbf{\emph{Acknowledgments}}
This work was supported by the National Natural Science Foundation of
China (Grant Nos.10965006, 11165014), and in part by the Project of Knowledge Innovation Program (PKIP) of Chinese Academy of Sciences, Grant No. KJCX2.YW.W10.

\end{document}